\documentclass{article}
\usepackage[T1]{fontenc} % add special characters (e.g., umlaute)
\usepackage[utf8]{inputenc} % set utf-8 as default input encoding
\usepackage{ismir,amsmath,cite,url}
\usepackage{graphicx}
\usepackage{color}
\usepackage{amsmath}
\usepackage{graphicx}
\usepackage{booktabs}
\usepackage{booktabs}
\usepackage{multirow}
\usepackage{siunitx}

\usepackage{microtype}

\usepackage{microtype}

\usepackage{soul}

\usepackage{lineno}
\title{Continual Learning for Singing Voice Separation with Human in the Loop Adaptation}

\multauthor
{Ankur Gupta* \hspace{1cm} Anshul Rai* \hspace{1cm} Archit Bansal* \hspace{1cm}  Vipul Arora \hspace{1cm}}
{Indian Institute of Technology, Kanpur, India\\
{\tt\small ankugupt@iitk.ac.in, anshulra@iitk.ac.in, architb@iitk.ac.in, vipular@iitk.ac.in}\\
{\footnotesize *Equal Contribution}}

\def\authorname{A. Gupta, A. Rai, A. Bansal and V. Arora}

\begin{document}

\maketitle
\begin{abstract}
% The abstract should be placed at the top left column and should contain about 150-200 words.

    Deep learning-based works for singing voice separation have performed exceptionally well in the recent past. However, most of these works do not focus on allowing users to interact with the model to improve performance. This can be crucial when deploying the model in real-world scenarios where music tracks can vary from the original training data in both genre and instruments. In this paper, we present a deep learning-based interactive continual learning framework for singing voice separation that allows users to fine-tune the vocal separation model to conform it to new target songs. We use a U-Net-based base model architecture that produces a mask for separating vocals from the spectrogram, followed by a human-in-the-loop task where the user provides feedback by marking a few false positives, i.e., regions in the extracted vocals that should have been silence. We propose two continual learning algorithms. Experiments substantiate the improvement in singing voice separation performance by the proposed algorithms over the base model in intra-dataset and inter-dataset settings. %and the fine-tuned model shows significant improvement in performance. %We got a mean SDR gain of 40.7\% on the MUSDB18 test dataset relative to the base performance.
\end{abstract}
\section{Introduction}\label{sec:introduction}

Singing voice or vocal is one the most important feature of music and is fundamental to human expression. Separation of vocals from polyphonic music tracks has been a research problem for quite some time as separated singing voice finds various applications in tasks like: singer identification\cite{Mesaros2007SingerII}, lyric transcription \cite{10.5555/2907324.2907380}, timer based artist clustering \cite{5410057}, beats and notes tracking \cite{inproceedings}, lyrics synchronization \cite{10.1145/2702123.2702140}, etc. 

%grouping artists by the timbre of their voice \cite{5410057},

The singing voice separation task has been approached using various signal processing methods such as \cite{8169990,10.1007/978-3-642-15995-4_18,article} in the past. In recent years, researchers have significantly improved the performance of vocal separation models using deep learning techniques \cite{huang2014singing, fan_jang_lu_2016, andreas_jansson_2017_1414934, lin_balamurali_koh_lui_herremans_2018, stoller2018waveunet, luo2018tasnet}. Even though these deep learning-based approaches provide state-of-the-art performance, they do not provide the user any way to conform the model to a particular song or genre and suffer greatly when new instruments or genres of music are used.
%\AR{[]}
This limits their scope of application to only songs which are similar to the training data (\textit{e.g.} if accompaniment of a test track has flute, but no song in the training data has flute, the model performance might suffer). Along with this, in the field of music, the data is continually evolving, and it is not feasible to continually create a new dataset to retrain the model from scratch. 

\begin{figure}
 \centering{
 \includegraphics[width=\columnwidth]{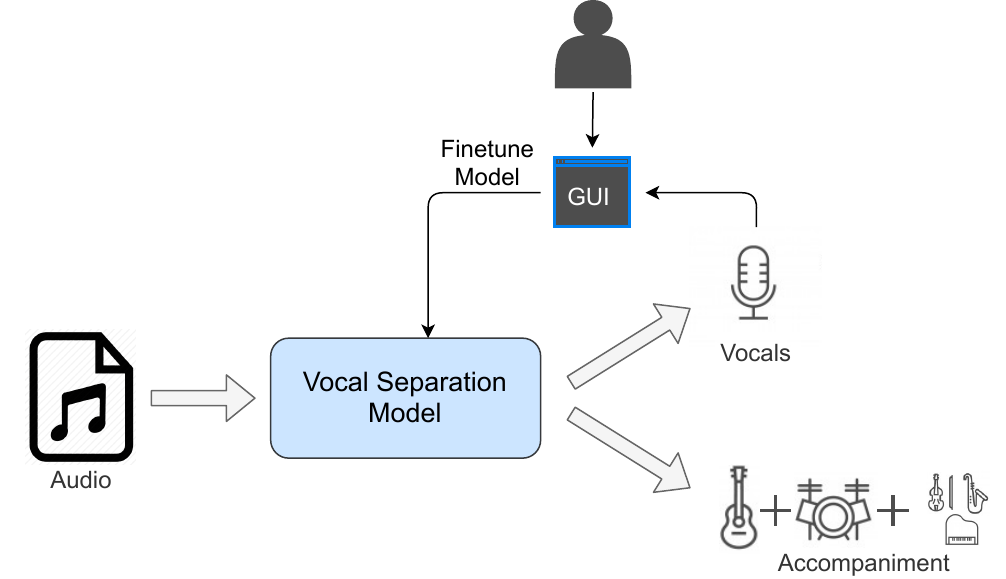}}
 \caption{\small{Pictorial view of problem statement}}
 \label{fig:ps}
\end{figure}

In this work, we have presented a user-friendly human-in-the-loop framework for continual learning to conform a U-Net-based deep learning model to target songs/genres while maintaining the generality of the model and avoiding overfitting as well as catastrophic forgetting. As shown in Fig. \ref{fig:ps}, the base U-net model is used to predict the vocal and accompaniment stems for the target songs. The model is then adapted using a human-in-the-loop approach, where first the human users mark regions in the predicted vocals where the model had incorrectly predicted vocals. Then this human feedback is used to update the model parameters, thereby increasing the model performance on the target songs. We have proposed two approaches for these fine-tuning steps that take special care towards increasing the model's generalisability.
 
To show the robustness of our approach, we have experimented under the intra-dataset condition where the test data belongs to the same dataset as the original train data as well as under inter-dataset condition where the model is initially trained on a different dataset and is adapted to the new dataset by correcting its mistakes by a human user on examples from the new-dataset. We also show that using our methods for singing voice separation increases performance not only for the songs corrected in the human-in-the-loop task but also for new songs that were previously unseen. The performance has been evaluated on different metrics which are compared on the basis of subjective separation quality.

Our contribution can be summarized as follows:
\begin{itemize}
    \item We propose a human-in-the-loop framework for user interaction with minimal effort and minimal music knowledge from the user's end.
    \item We present two continual learning techniques for model fine-tuning with the aim to minimize overfitting and catastrophic forgetting as well as to increase model generalisability while adapting to a new test song. 
    \item We show that for voice separation, continual learning problems mean of SDR is a better measure than widely used median SDR.
\end{itemize}

The remaining part of the paper is organized as follows. In Section \ref{sec:prev} we have discussed previous work in the fields of source separation, human-in-the-loop \& continual learning. In Section \ref{sec:pm} we have introduced and explained our proposed methods. This is followed by details of our experimentation and an analysis of the results in Section \ref{sec:Experimentation}. We have concluded with the closing remarks in Section \ref{sec:conc}.  We have also released
the source code along with the paper\footnote{link will be updated after the anonymity period}

% This template includes all the information about formatting manuscripts for the ISMIR \conferenceyear\ Conference.
% Please follow these guidelines to give the final proceedings a uniform look.
% Most of the required formatting is achieved automatically by using the 
% supplied
% style file (\LaTeX) or template (Word).
% If you have any questions, please contact the Program Committee (\texttt{ismir\conferenceyear-papers@ismir.net}).
% This template can be downloaded from the ISMIR \conferenceyear\ web site (\texttt{http://ismir\conferenceyear.ismir.net}).
%

\section{Previous Work}
\label{sec:prev}
Many deep learning-based approaches with great performance for singing voice separation have been explored in the recent past. Approaches like \cite{huang2014singing, fan_jang_lu_2016} use deep RNN architectures on the magnitude plot of the short-time fourier transform (STFT).  \cite{andreas_jansson_2017_1414934} introduces U-net based approach to the music domain for singing voice separation. \cite{lin_balamurali_koh_lui_herremans_2018} uses deep CNN architecture to estimate the source label for each (T-F) bin in the spectrogram. \cite{stoller2018waveunet,luo2018tasnet} present an approach for source separation directly in the time domain, thereby making use of time-domain signals rather than frequency spectrogram. In this paper, we present our method using U-net architecture, but our method is independent of the exact architecture used and can be applied to other architectures too.

Most of the deep learning-based approaches for singing voice separation focus only on increasing the model performance on benchmark metrics and don't focus on conforming the model to particular songs and/or genre in a continual learning setting. Previous work for adapting to particular songs have been through human feedback approaches that require either the human to do spectrogram-based mask editing \cite{10.1145/2556288.2557253}, or provide feedback on the fundamental notes/frequency of the singing voice in the song  \cite{nakano_koyama_hamasaki_goto_2020}. These approaches hence assume a minimum level of musical understanding in the user and are highly time-consuming. Moreover, the performance of these methods involves overfitting the model to the particular song making their lifelong learning aspect weak. Approaches like \cite{5946389} which use time activation of a particular instrument as annotations from the user don't use deep learning methods. In this paper, we propose a simple human-in-the-loop task that requires minimal music knowledge and can be applied to a large group of deep learning models.  

Learning from an infinite stream of data is studied under the continual learning paradigm. \cite{Delange_2021} surveyed the different approaches that have been proposed for tackling this task and divided them into three categories: replay-based, regularization-based, and parameter isolation methods. Replay-based methods \cite{rolnick2019experience, rebuffi2017icarl, isele2018selective} rely on storing exemplars from the original training data or generating data points similar to the training data using generative models to prevent forgetting and overfitting. Regularization-based methods like \cite{mazumder2021fewshot} have proposed freezing important weights of the initial model and regularizing trainable weights during finetuning to deal with catastrophic forgetting. Parameter isolation methods dedicate separate parameters to each task in order to prevent catastrophic forgetting, using techniques such as network pruning \cite{mallya2018packnet}, introducing new branches to the model\cite{rusu2016progressive} and masking \cite{mallya2018piggyback}. In this paper, we adapt replay-based continual learning algorithms to use for model adaptation on a stream of user annotated data. 

%https://hal.inria.fr/inria-00564851/document
%  We develop an approach based on U-net model for singing voice separation which just requires the user to identify the regions which don't contain any vocals and label them. This method is useful in cases when the new song differs in genre or instruments when compared to our training data. It provides the user an ability to fine-tune the model even in the lack of ground truth labels. Further, the learning from these labels is transferable to other similar songs and hence improves the model performance in the long run.

%\AR{continual learning methods need to be discussed}

\section{Proposed Method}%\label{sec:page_size}
\label{sec:pm}
\subsection{Problem Setting}
The problem of singing voice separation has been formally defined as given a time-domain polyphonic music signal $M(t)$, separate $M(t)$ into its constituent signals $V(t)$ and $A(t)$ such that they have minimal interference. Here $V(t)$ and $A(t)$ denotes the separated vocals and accompaniment of the original music signal.

Given a dataset D, we split it into 3 components $\mathbf{D_{train}}$, $\mathbf{D_{HITL}}$ and $\mathbf{D_{test}}$, where $\mathbf{D_{train}}$  is used for training the base model, $\mathbf{D_{HITL}}$ is used for human in the loop adaptation, while $\mathbf{D_{test}}$ is the unseen set used for testing the model as in usual machine learning cases. We first train a vocal separation model $\mathbf{B}$ using $\mathbf{D_{train}}$ which is our base model. Next, we define the continual learning human-in-the-loop task as follows: Given the model $\mathbf{B}$, for each target song $s$, ( $s \in \mathbf{D_{HITL}}$ ) modify the parameters of $\mathbf{B}$ based on human annotations on $s$, such that its performance on $s$ and songs similar to $s$ increases. Finally, we evaluate the performance of our model on the $\mathbf{D_{test}}$ set to test that our model does not suffer from overfitting or catastrophic forgetting.
% while avoiding forgetting what the model has already learnt (catastrophic forgetting) instead of overfitting to $s$.  %We wish to keep the human labelling task simple, and something which even a person not familiar with music theory can easily do. 

\begin{figure*}[t]
 \centering{
 \includegraphics[width=\textwidth]{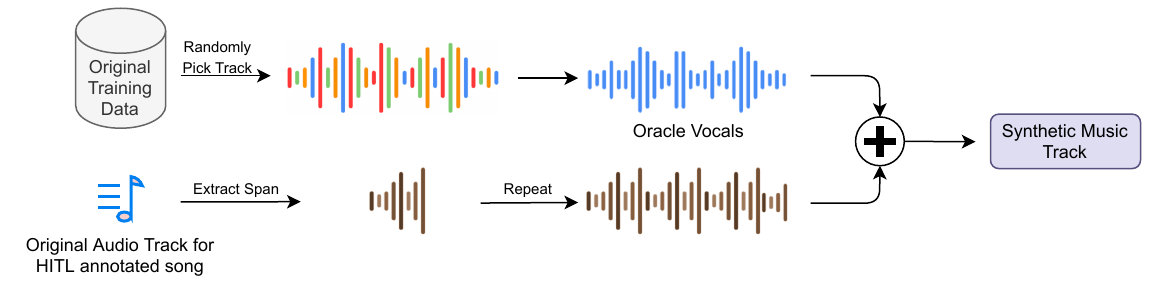}}
 \caption{Pipeline for creating synthetic music tracks}
 \label{fig:example2}
\end{figure*}

\subsection{Method Overview}

For our base model, we use a U-Net-based model similar to \cite{Jansson2017SingingVS}. The U-Net architecture consists of an encoder-decoder network made up of convolutional layers with skip connections between the encoder and decoder layers which allow retention of any information lost during encoding. The base model was trained on the $\mathbf{D_{train}}$ split of the MUSDB-18 dataset  \cite{musdb18}, taking data points $(X,Y)$ from each music signal $M(t) \in \mathbf{D_{train}}$ , where $X$ is the corresponding spectrogram sliced from the spectrogram of $M(t)$ and $Y$ is the target mask. We use L1 loss for the vocal masks and optimize with Adam optimizer, using a learning rate of $10^{-4}$.

%train a base model for this task using train split of the MUSDB-18 dataset, with a learning rate of $10^{-4}$.

%Only 10 frames are taken from each song to keep each songs contribution to the model constant irrespective of the song length, similar to [].
For the human-in-the-loop adaptation, we have formulated a human annotation task where the human has to mark segments $t_i$ with the start and end timestamps, $t^{start}_{i}$ and $t^{end}_{i}$, %in mixed (input) audio track $M(t)$, 
where the estimated vocal track $\Hat{V}(t)$ contains audio while it should have been silent. Here, $i$ is the segment index.
%the time frame ($t_{start_{i}}$,$t_{end_{i}}$) in $M(t)$ should not contain any vocals, but the model predicted vocals $\Hat{V}$ contain residual noise instead of silence in that time frame. 
The annotation task is kept simple, and it is formulated such that the annotator need not have any specialized music skills. During our experiments, we have used a simple heuristic of keeping only the marked regions that are longer than $6$ seconds in length to ensure uniformity. We have also developed a GUI interface to make it easier for users to mark the timestamps. 

 We use batch gradient-based fine tuning for adapting the model. L1 loss is used on the vocal mask along with Adam optimizer, similar to how we had trained our base model, but with 10 times lower learning rate. We fine-tune the model for a fixed number of epochs.

\subsubsection{Adaptation with Zero Vocal Targets}
In the regions $t_i$ as there are no vocals to extract, the ideal vocal masks for these regions should identically be equal to zero. Therefore the target to the model $\mathbf{B}$ for these regions are zero masks.
For model adaptation in this method, we use only the segments $t_i$ for training as we know their target masks.

% \VA{Explain the Fine tuning method}

%\textbf{A. Simple Adaptation:} 

%by taking data points $(X,Y)$ for each timeframe, where $X$ is the corresponding spectrogram sliced from the spectrogram of $M(t)$ and and $Y$ is the target mask set identically equal to zero,  
%This is because the learning task becomes too simple with the model solely learning to predict only zeros. 

%  We observed that if we perform direct fine-tuning using the human annotations by taking $(X,Y)$ where $X$ is the spectrogram of $M(t)$ in the region of no vocals and $Y$ is the target mask set identically equal to zero our model quickly over fit to these areas. This makes sense as the learning which would happen would be to identically predict zero instead of the correct vocal mask %{\AG{ we can add something more here}.
%  To avoid this along with, catastrophic forgetting and overfitting problem that occur in continual learning scenarios we explore two methods: A. Replay based method B. Artificial track creation. 

\begin{figure}[h]
 \centering{
 \includegraphics[width=\columnwidth]{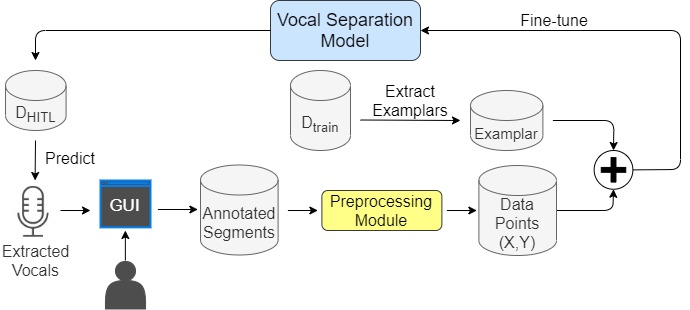}}
 \caption{\small{Replay based continual learning}}
 \label{fig:replayimage}
\end{figure}

% \subsubsection{Replay Based Method} 
% \label{subsec:Replay Based Method}
%In this approach, the user marks the regions that the base model gets wrong and predicts vocals when there should have been silence.
% To avoid this, along with the issues of catastrophic forgetting and overfitting which occur in continual learning scenarios, we use replay-based continual learning approach \VA{cite}.
We observed that if we perform direct fine-tuning using the human annotations of $\mathbf{D_{HITL}}$, our model quickly overfits these areas. This makes sense as the learning would be to identically predict zero instead of the correct vocal mask. To ensure that the model does not overfit to the new data and does not forget the knowledge it gained from previous examples, replay based continual learning method, which relies on storing a subset of the previous training data ($\mathbf{D_{train}}$), is used while fine-tuning the model. As shown in Fig.~\ref{fig:replayimage}, we update the model on exemplars from the original training data ($\mathbf{D_{train}}$) in conjunction with the data points extracted from the regions marked by the user during the HITL task on $\mathbf{D_{HITL}}$. %For updating the model weights, we use L1 loss on vocal mask with Adam optimizer similar to how we had trained our base model, but with a 10 times smaller learning rate. 

\subsubsection{Adaptation with Synthetic Track}
In the above method, %performed well, 
the new data points given to the model from the user annotated regions are expected to be silent. These are not informative enough to have a good learning. In order to improve the quality of those new data points, we decided to combine the human-in-the-loop annotations and the original training data ($\mathbf{D_{train}}$) to form synthetic music tracks which provide more informative data points. 
\par As shown in Fig.~\ref{fig:example2}, we create synthetic tracks by superimposing the oracle vocals of songs from $\mathbf{D_{train}}$ with the user annotated time frames from $\mathbf{D_{HITL}}$. This is achieved by extracting the annotated time segments $t_i$ from $M(t) \in \mathbf{D_{HITL}}$ and repeating it to form the artificial accompaniment for the synthetic track. A track is randomly chosen from $\mathbf{D_{train}}$, and its oracle vocals are superimposed on the artificial accompaniment. During the finetuning stage, we extract data points $(X, Y)$ from these synthetic tracks, where X is a slice from the spectrogram of the artificial track and Y is the corresponding slice from the oracle vocals used for creating that track. As before, the model is updated in conjunction with these data points and the stored exemplars from $\mathbf{D_{train}}$, using L1 loss and Adam optimizer with learning rate $10^{-5}$.

%\begin{figure*}[!ht]
%\includegraphics[width=\textwidth]{figs/fig1.png}
%\end{figure*}

% The proceedings will be printed on
%  \underline{portrait A4-size paper} \underline{(21.0cm x 29.7cm)}.
% All material on each page should fit within a rectangle of 17.2cm x 25.2cm,
% centered on the page, beginning 2.0cm
% from the top of the page and ending with 2.5cm from the bottom.
% The left and right margins should be 1.9cm.
% The text should be in two 8.2cm columns with a 0.8cm gutter.
% All text must be in a two-column format.
% Text must be fully justified.

\section{Experiments and Results}\label{sec:Experimentation}

\begin{figure*}[h]
\centering
\includegraphics[width=2\columnwidth]{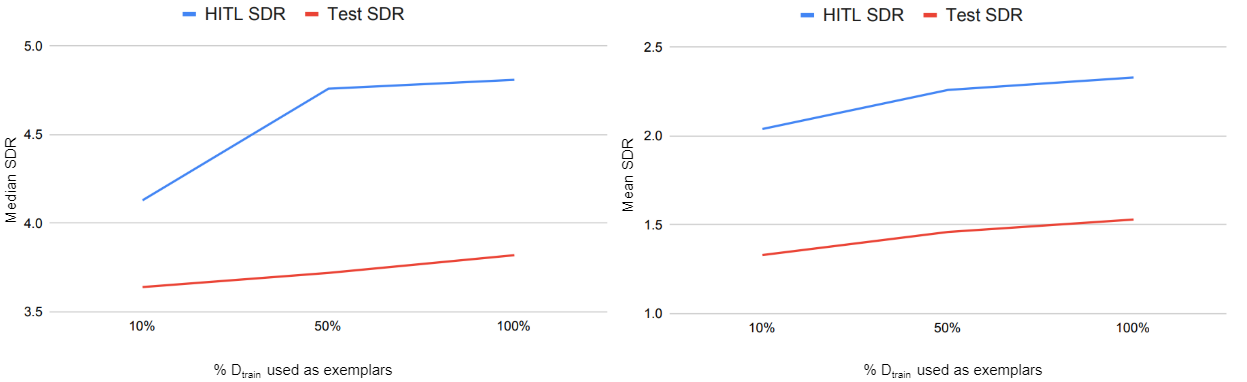}
 \caption{\small{Comparison of model performance ( Mean \& Median SDR ) with 10\%, 50\% and 100\% of train data as exemplar}}
 \label{fig:Vocal Mean SDR _ MUSDB18 Test Set A}
\end{figure*}

For the methods proposed in this paper, we have conducted extensive experiments for both inter-dataset and intra-dataset settings. The results obtained have been tabulated and studied in this section. 

\textbf{Dataset}: We have experimented with the standard benchmark MUSDB-18 dataset to test the performance of our proposed approaches.
From the train split, consisting of 100 songs, we have used 90 songs to form $\mathbf{D_{train}}$ and the remaining 10 songs to form $\mathbf{D_{HITL}}$. The test split, which consists of 50 songs, is used as $\mathbf{D_{test}}$. %to compute results that are a measure of our models generalisability after the HITL task and hence a measure of continual learning ability of our approach.
%\AR{list the 10 songs}

To simulate inter-dataset conditions, where the intention is to adapt the model to a new kind of previously unseen data without the knowledge of true stems, we perform human-in-the-loop adaption using a subset of the CC-Mixter dataset\cite{6842708} and test the performance on the leftover test set in CC-Mixter dataset. Table \ref{tab:dataset} contains the number of songs in different sets. The MUSDB-18 dataset consisted of vocals, drums, bass, and the rest of the accompaniment stems, which we combined to create two stems, vocals, and accompaniment. The CC-Mixter contained only vocals and accompaniment stems. For all occurrences in tables, we denote the MUSDB-18 dataset by MUS and the CC-Mixter dataset by CCM.

\begin{table}[h]
\small
\centering
\begin{tabular}{ cccc } 
 \toprule
 \bfseries Dataset & $\mathbf{D_{train}}$ & $\mathbf{D_{HITL}}$ & $\mathbf{D_{test}}$ \\
 \midrule
 MUS & 90 & 10 & 50  \\ 
 CCM & - & 30 & 20 \\ 
 \bottomrule
\end{tabular}
\vspace{-0.5em}
\caption{\small{Distribution of the number of songs in different split ( train, HITL \& test ) of the datasets}}
\label{tab:dataset}
\end{table}

%. and performance is compared for different human-in-the-loop methods and the base method. We have also performed human-in-the-loop adaptation on x songs from the CC-mixter dataset[], this is referred to as the CC-mixter HITL set. Finally performance of the base model and the fine-tuned model has been evaluated on x songs of the CC-mixter dataset, hereby referred as the CC-mixter test set.

\textbf{Pre-processing:}  $M(t)$ is made monaural and re-sampled to $22050$ Hz before its short-time Fourier transform (STFT) is computed\footnote{computed using librosa\cite{brian_mcfee_2020_3955228} library} with a window size of $2048$ and a stride of $512$. The magnitude of these STFT spectrograms is divided into windows of shape $(1,1024,512)$ [channel, frequency, time] and used as input for our U-net model, which provides us with a mask for extracting vocals from the mixture spectrogram. Using spectrogram of $M(t)$ and this vocal mask from the model, we can compute our estimated $V(t)$ using inverse short-time Fourier transform (iSTFT).

\textbf{Evaluation Metric:} As proposed by \cite{1643671}, we use the Signal Distortion Ratio (SDR) for evaluation of predicted vocal \& accompaniment stem\footnote{We used the pre-existing library for SDR calculation: https://sigsep.github.io/sigsep-mus-eval/}. We aggregate the SDR using a mean over all 1-second frames to keep one single metric per song. While many previous works report the median SDR of each song, we use mean SDR as we found it's a better measure to evaluate the performance ( explained in section \ref{subsec:Mean SDR over Median SDR} ). However, to make our work comparable with other standard works, we have also reported the median SDR wherever necessary. Due to similar reasons when dealing with multiple songs, we report the mean of SDR computed across all songs, whereas many papers report the median of the song-wise SDR across the songs.                                                                                                       

\subsection{Adaptation with Zero Vocal Targets}\label{subsec:Results and analysis}
%\AR{insert Graphs}
For our replay-based HITL adaptation with zero vocal targets, we had to identify the number of exemplars needed during finetuning for the best results. We observe that the performance of the finetuned models increased on both the median and mean SDR  metrics with the number of exemplars kept during replay as seen from figure \ref{fig:Vocal Mean SDR _ MUSDB18 Test Set A}. As a result, for all the models adapted using the zero vocal targets algorithm, we keep 100$\%$ of the training data as exemplars and finetune the model for a single epoch. Table \ref{tab:replaymusdb18_1} compares the performance of the base vocal separation model (U-Net(MUS)) trained on $\mathbf{D_{train}}$ of the MUSDB-18 dataset and the model obtained after HITL adaptation with zero vocal targets (U-Net(MUS) + HITL(MUS)) on the $\mathbf{D_{HITL}}$ set of the MUSDB-18 dataset.  The results show an improvement in the mean SDR for both the $\mathbf{D_{HITL}}$ and $\mathbf{D_{test}}$ sets of the MUSDB-18 dataset. This indicates that the model finetuned on just the $\mathbf{D_{HITL}}$ set did not suffer from overfitting and was generalizable such that it is able to learn to improve non-vocal regions in new songs as well. However, the results in Table \ref{tab:replaymusdb18_1} show a slight decrease in the median SDR over both the sets $\mathbf{D_{HITL}}$ and $\mathbf{D_{test}}$. This is not a big issue because, as we go on to show in subsequent sections, mean SDR is a better metric for evaluating the overall quality of the separated vocals, provided the median SDR is within a reasonable range of the base model.
 
\begin{table}[h]
\small
\centering
\begin{tabular}{ lSS } 
 \toprule
  \bfseries Model & $\mathbf{D_{HITL}}$ & $\mathbf{D_{test}}$ \\
 \midrule
  \multirow{2}{*}{} & \multicolumn{2}{c}{Mean} \\
 U-Net(MUS)  & 1.38 & 1.13  \\ 
 U-Net(MUS) + HITL(MUS) & 2.32 & 1.53 \\ 
 %\vspace{+0.5em}
  \multirow{2}{*}{} & \multicolumn{2}{c}{Median} \\
 U-Net(MUS)  & 5.02 & 3.87  \\ 
 U-Net(MUS) + HITL(MUS) & 4.81 & 3.82 \\ 
 \bottomrule
\end{tabular}
\vspace{-0.5em}
\caption{\small{SDR score on $\mathbf{D_{HITL}}$ and $\mathbf{D_{test}}$ set of MUS}}
\label{tab:replaymusdb18_1}
\end{table}

\begin{table}[h]
\small
\centering
\begin{tabular}{ lSS } 
 \toprule
  \bfseries Model & $\mathbf{D_{HITL}}$ & $\mathbf{D_{test}}$ \\
 \midrule
  \multirow{2}{*}{} & \multicolumn{2}{c}{Mean} \\
 U-Net(MUS)  & -6.29 & -2.35  \\ 
 U-Net(MUS) + HITL(CCM) & -3.37 & -1.02 \\ 
 %\vspace{+0.5em}
  \multirow{2}{*}{} & \multicolumn{2}{c}{Median} \\
 U-Net(MUS)  & -0.37 & 4.60  \\ 
 U-Net(MUS) + HITL(CCM) & 0.27 & 2.64 \\ 
 \bottomrule
\end{tabular}
\vspace{-0.5em}
\caption{\small{SDR score on $\mathbf{D_{HITL}}$ and $\mathbf{D_{test}}$ set of CCM}}
\label{tab:replayccm_1}
\end{table}

To verify the adaptability of our approach, we perform inter-dataset validation using the CC-Mixter as our target dataset. We used the pre-trained base model from the previous experiment and adapted it to this new dataset using the HITL set ( $\mathbf{D_{HITL}}$ ) of the CC-Mixter dataset over a single epoch. We can see from Table \ref{tab:replayccm_1} that the songs of the CC-mixter dataset show a significant increment in mean SDR for the HITL set from -6.29 to -3.37 and for the Test set ( $\mathbf{D_{test}}$ ) from -2.35 to -1.02. This increment in performance once again exhibits the quick adaptability of the model to the $\mathbf{D_{HITL}}$ set by the HITL process as well as a general improvement in the ability of the model to suppress residual noise in songs of both the $\mathbf{D_{HITL}}$ and $\mathbf{D_{test}}$ sets. The results in Table \ref{tab:replayccm_1} however, reveal an issue with the zero vocal target adaptation methodology. As the new HITL data points do not have any information about vocals but rather only about the residual noise which the base model failed to remove, the finetuned model, after finetuning over a larger $\mathbf{D_{HITL}}$ set of 30 songs, saw a decrement in the performance on vocals as seen with the decrease in median SDR of the $\mathbf{D_{test}}$ set. This shows that our hypothesis about the zero vocal targets not containing enough information was true.

\subsection{Adaptation with Synthetic Track} \label{subsec: Replay with Artificial Track}
Our proposed method of adaptation using synthetic music tracks, as elaborated in section 3 shows an improvement in the both mean and median SDR of the $\mathbf{D_{HITL}}$ and $\mathbf{D_{test}}$ of MUSDB-18 dataset in comparison to the zero vocal target optimization scheme ( Table \ref{tab:replaymusdb18_2} ). The model was trained such that for each synthetic music track, there were z batches, with each batch containing x data points from the synthetic music track and y data points randomly chosen from the training data set. After performing hyperparameter tuning using grid search, we were able to set the values of each of the hyperparameters to x=1, z=1, and y=15 for all of our reported results,
%\AR{Fig 6} shows the dependence of z, x, and y on the performance.
Furthermore, the model finetuned using this strategy was able to perform better than our previous method using just 20$\%$ of the original training data ( $\mathbf{D_{train}}$ ), while in comparison, the zero target vocal adaptation model used 100$\%$ of the training data.

%\begin{figure}[h]
% \centering{
 %\includegraphics[width=\columnwidth]{figs/Vocal Mean SDR _ MUSDB18 Test Set A (Batch Size=16).pdf}}
%\caption{\small{Figure captions should be placed below the figure}}
% \label{fig:Vocal Mean SDR _ MUSDB18 Test Set A}
%\end{figure}

\begin{table}[h]
\small
\centering
\begin{tabular}{ lSS } 
 \toprule
  \bfseries Model & $\mathbf{D_{HITL}}$ & $\mathbf{D_{test}}$ \\
 \midrule
  \multirow{2}{*}{} & \multicolumn{2}{c}{Mean} \\
 U-Net(MUS)  & 1.38 & 1.13 \\ 
 U-Net(MUS) + HITL(MUS) & 2.45 & 1.59 \\ 
 %\vspace{+0.5em}
  \multirow{2}{*}{} & \multicolumn{2}{c}{Median} \\
 U-Net(MUS)  & 5.02 & 3.87  \\ 
 U-Net(MUS) + HITL(MUS) & 4.89 & 3.93 \\ 
 \bottomrule
\end{tabular}
\vspace{-0.5em}
\caption{\small{SDR score on $\mathbf{D_{HITL}}$ and $\mathbf{D_{test}}$ set of MUS}}
\label{tab:replaymusdb18_2}
\end{table}

\begin{table}[h]
\small
\centering
\begin{tabular}{ lSS } 
 \toprule
  \bfseries Model & $\mathbf{D_{HITL}}$ & $\mathbf{D_{test}}$ \\
 \midrule
  \multirow{2}{*}{} & \multicolumn{2}{c}{Mean} \\
 U-Net(MUS)  & -6.29 & -2.35   \\ 
 U-Net(MUS) + HITL(CCM) & -4.32 & -1.36 \\ 
 %\vspace{+0.5em}
  \multirow{2}{*}{} & \multicolumn{2}{c}{Median} \\
 U-Net(MUS)   & -0.37 & 4.60  \\ 
 U-Net(MUS) + HITL(CCM) & 0.83 & 4.46 \\ 
 \bottomrule
\end{tabular}
\vspace{-0.5em}
\caption{\small{SDR score on $\mathbf{D_{HITL}}$ and $\mathbf{D_{test}}$ set of CCM}}
\label{tab:replayccm_2}
\end{table}

Similar to the previous part, we also analyzed the model performance on inter-dataset conditions. Table \ref{tab:replayccm_2} shows our performance on the HITL ( $\mathbf{D_{HITL}}$ ) and test set ( $\mathbf{D_{test}}$ ) of the CC-Mixter dataset. We see that the model is once again able to improve performance on the mean SDR metric for both the songs on which the user-provided annotations ( $\mathbf{D_{HITL}}$ ) as well as on unseen songs ( $\mathbf{D_{test}}$ ). Furthermore, we can see from Table \ref{tab:replayccm_2} that the issue we faced regarding decrement in the median SDR due to zero vocal targets has been completely eradicated. This once again confirms our hypothesis that the use of synthetic tracks to furnish data points for our model adaptation provides much more information to the model regarding both residual noise and vocal separation. The results in Table \ref{tab:replaymusdb18_2} and Table \ref{tab:replayccm_2} show that using artificial tracks, we get the best of both worlds - better performance with minimal data requirement.

\subsection{Mean SDR over Median SDR} \label{subsec:Mean SDR over Median SDR}
The SDR for a song is estimated using a statistical measure (mean/median) after determining SDR for each 1-second frame of the song. Though some previous works on music source separation have reported their result using median, they did not involved continual learning. For human-in-the-loop continual learning problems like ours, we saw mean to be a better measure of performance than the median. 

\begin{figure}[h]
 \centering{
 \includegraphics[width=\columnwidth]{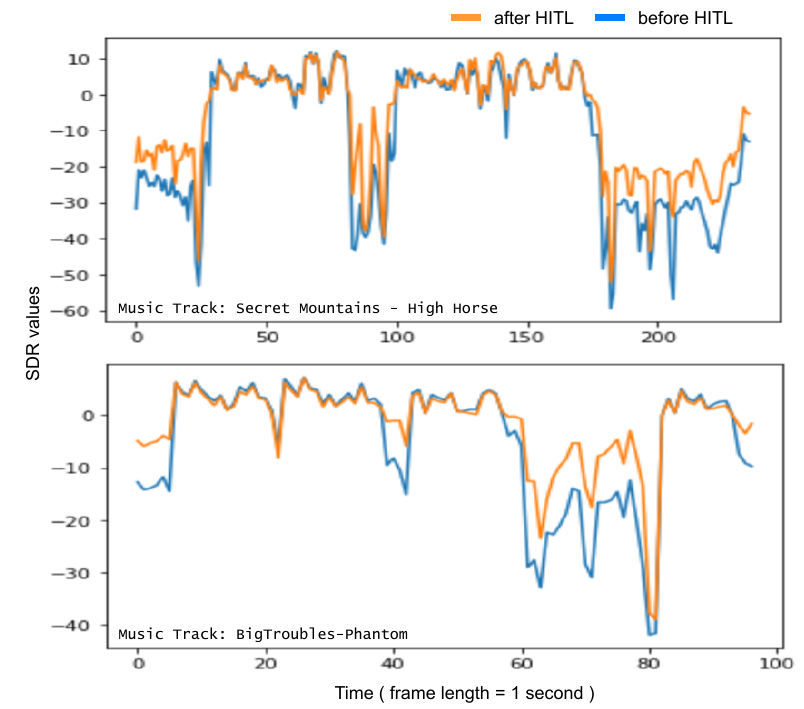}}
 \caption{\small{Comparison of frame wise SDR of songs before and after HITL}}
 \vspace{-0.5em}
 \label{fig:song}
\end{figure}

Fig \ref{fig:song} shows the graph for SDR (in y -axis) with time(in x-axis) for the two songs of the test set ( $\mathbf{D_{test}}$ ) of MUSDB18 dataset: Secret Mountains - High Horse and BigTroubles-Phantom before (blue) and after (orange) the HITL adaptation. 
%Only one block of the song was marked by the human annotator as silence region for the vocal during HITL.
We examine from the graph that SDR increases by a considerable margin in the regions with previously low SDR values. At the same time, in areas with already high SDR, performance is similar after HITL when compared to previous base performance. As we can see in fig 5 majority of data points are inclined towards this higher SDR region where the SDR remains almost unchanged, so the median value is not affected after HITL. However, since the mean considers all the frames equally, we see a considerable rise in the overall mean SDR score of the song after HITL. The main aim of the HITL task is to improve the performance of regions that perform poorly (usually low SDR regions) by giving feedback to the model about its mistakes which indeed is happening after performing HITL with our model. An increase in performance on low SDR regions leads to an overall improvement in sound quality when heard by human ears.%\footnote{relevant predicted stem files of both songs are provided in the supplementary material for reference} 

\subsection{Continual Learning} 
\begin{figure}[h]
 \centering{
 \includegraphics[width=\columnwidth]{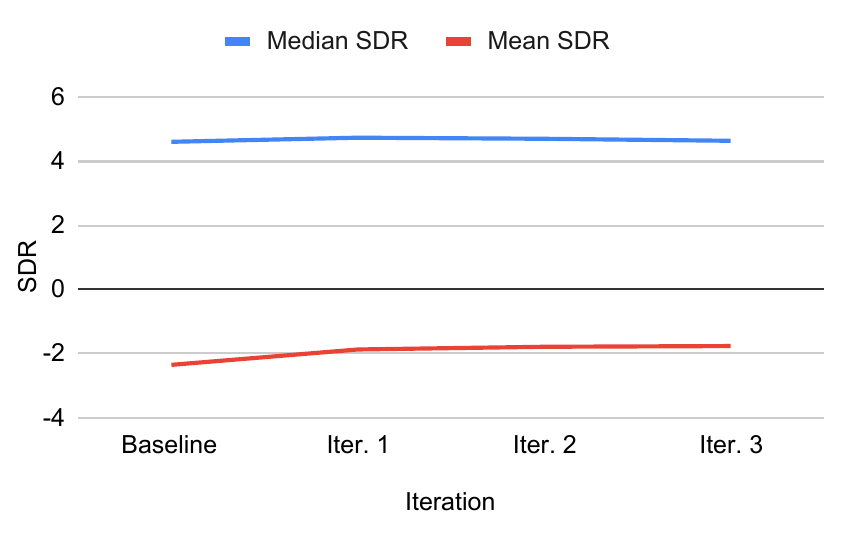}}
 \vspace{-0.5em}
 \caption{\small{Comparison of SDR after each iteration of HITL}}
 \label{fig:iteration}
\end{figure}

In order to show that our continual learning approach is free of overfitting, we iteratively performed the HITL task 3 times, each time with a HITL batch of 10 marked audio samples. We used the U-Net model trained with MUSDB18 as Base and performed HITL with HITL set ( $\mathbf{D_{HITL}}$ ) of CC-mixter dataset in a way similar to inter-dataset setting in the section \ref{subsec: Replay with Artificial Track}.  Figure \ref{fig:iteration} shows the mean and median SDR values pictorially after each iteration. Note that the SDR values have been evaluated on the complete CC-mixter test set ( $\mathbf{D_{test}}$ ) after each iteration.  We see that the median value remains almost constant, indicating that our model does not overfit when learning its mistakes on low SDR parts, and its performance remains intact in other parts. Moreover, we see that the mean value keeps increasing gradually with a significant gain after the first iteration itself, indicating that the model adapts to the new data quickly and keeps on improving with the addition of more examples, but doesn't overfit to HITL set songs.

For measuring catastrophic forgetting, we computed the SDR score on the test set of MUSDB-18 ( $\mathbf{D_{test}}$ ) after performing HITL fine-tuning on the HITL set of CC-mixter ( $\mathbf{D_{HITL}}$ ). We got a median SDR score of 3.75 and a mean SDR score of 1.76, which is well within the desired range.

\section{Conclusion and Future Work} 
\label{sec:conc}

In this paper, we present a framework to incorporate human feedback to improve the performance of vocal separation models on the existing dataset and adapt the current model to other kinds of the real-world dataset without requiring actual vocals stems. 
We presented two adaptation algorithms - one using zero vocal targets and the other using synthetic tracks - the latter performed better than the former in our experiments and also used lesser exemplars from the $\mathbf{D_{train}}$ set, thereby reducing storage requirements and decreasing adaptation time.
% We showed that with additional synthetically generated tracks during fine-tuning, we get an appreciable rise in SDR performance compared to that from other methods. 
We showed that our final continual learning approach minimizes overfitting and catastrophic forgetting and maximizes generality. We also showed that reporting mean SDR is a better evaluation criterion than median SDR in human-in-the-loop source separation tasks. Also, approaches in this paper are independent of the initial choice of the deep learning model. They hence could be used in conjunction with other network types with a little bit of fine-tuning for hyperparameters for best performance. 
%\AR{poore paper mein vocal ya singing voice karna hai kya}

Though this paper currently focuses on separating the vocals from the music signal, future works can explore extending this to other stems with adequate modifications to the human-in-the-loop task. Recently there have also been encouraging works in the field of regularization-based continual learning in the field of computer vision, and there are opportunities for future works to incorporate these algorithms in the proposed human-in-the-loop framework. 

\bibliography{ISMIRtemplate}

% For non bibtex users:
%\begin{thebibliography}{citations}
% \bibitem{Author:17}
% E.~Author and B.~Authour, ``The title of the conference paper,'' in {\em Proc.
% of the Int. Society for Music Information Retrieval Conf.}, (Suzhou, China),
% pp.~111--117, 2017.
%
% \bibitem{Someone:10}
% A.~Someone, B.~Someone, and C.~Someone, ``The title of the journal paper,''
%  {\em Journal of New Music Research}, vol.~A, pp.~111--222, September 2010.
%
% \bibitem{Person:20}
% O.~Person, {\em Title of the Book}.
% \newblock Montr\'{e}al, Canada: McGill-Queen's University Press, 2021.
%
% \bibitem{Person:09}
% F.~Person and S.~Person, ``Title of a chapter this book,'' in {\em A Book
% Containing Delightful Chapters} (A.~G. Editor, ed.), pp.~58--102, Tokyo,
% Japan: The Publisher, 2009.
%
%
%\end{thebibliography}

\end{document}